\documentclass[vecphys]{article}
\usepackage{amsmath,amssymb,mathrsfs,wick,hyperref}

\newcommand{\bx}{{\boldsymbol x}}

\newcommand{\ba}{{\boldsymbol a}}

\newcommand{\ux}{{\underline x}}
\newcommand{\uy}{{\underline y}}
\newcommand{\ttimes}{\tilde{\times}}
\newcommand{\I}{\iota}

\begin{document}

\title{DFR Perturbative Quantum Field theory on Quantum Space Time, and Wick 
Reduction} 
\author{Gherardo Piacitelli\thanks{Research 
partially supported by MIUR and GNAMPA-INdAM}\\
\tt{http://www.piacitelli.org}}
\date{Submitted on March 20, 2005}
\maketitle

\begin{abstract}
We discuss the perturbative approach \emph{\`a la Dyson} to a quantum field 
theory with nonlocal self--interaction  
\(\boldsymbol{:}\!\phi\star\dotsm\star\phi\!\boldsymbol{:}\), according to 
Doplicher, Fredenhagen and Roberts (DFR). In particular, we show that
the Wick reduction of \emph{non locally} time--ordered products of Wick 
monomials can be performed as usual, and we discuss a very simple 
Dyson diagram. 
\end{abstract}

\noindent\emph{Dedicated to Jacques Bros on the occasion of his 
\(70^{\text{th}}\) birthday. }

\section{Introduction}
\label{sec_1}
During the $\text{XX}^{\text{th}}$ 
century, locality has been so valuable a principle in the 
development of high energy physics, that it is strongly encoded in our minds.
Sometimes we advocate locality even when it is not strictly necessary. 

I once heard Daniel Kastler telling this story, that a sixty or so pages long 
paper appeared, about a certain topic in commutative algebra; the astonishing
fact was that the only necessary change in order to generalize the results
of that paper to the non abelian case was the removal of every occurrence of 
the word ``abelian''. The conclusion of Kastler's tale was that commutativity
comes so natural to our mind, that sometimes even an expert might overlook 
the generality of some arguments. 

Something similar happens with locality. We will see that some aspects
and methods of the perturbation theory of a certain class of nonlocal 
theories are exactly the same as in the local case. 

Indeed, we will consider the nonlocal theory which naturally
arises when attempting a perturbative quantum field theory \emph{ \`a la} 
Dyson 
on the flat DFR quantum spacetime. It could have been easily recognized 
that the usual diagrammatic representation of the correction terms to the 
trivial scattering matrix has nothing to do with locality. The interesting 
point, however, is that it has \emph{not} been easy to recognize this, probably
because of some psychological obstruction of the kind mentioned 
by Daniel Kastler.

In the next section, we will give a short description of the machinery 
underlying the DFR model of a flat quantum spacetime, and of a 
quantum field theory built on top of it. We will take the occasion to 
clarify the relations between the original DFR notations and those which
are now current in the literature (see also the appendix).

In section \ref{sect_3}, we will describe how to derive the Dyson diagrams
for a (possibly) nonlocal perturbation theory using precisely the
same methods which were developed in the late 40's. 
An analogous discussion can be done for the
Feynman diagrams, see \cite{newrules} for details.

We will draw some conclusions in section \ref{sect_4}.

\section{DFR Quantum Spacetime, and All That}

\subsection{The Underlying Philosophy}

In their seminal paper \cite{dfr}, Doplicher Fredenhagen and Roberts 
proposed to derive a simple model of spacetime coordinates quantization, 
stemming from first principles endowed with an operational meaning.

Indeed, the idea of quantizing the coordinates was quite old \cite{snyder}.
The idea that non commuting coordinates would produce a coarse grained
spacetime also was far from new. However, the spirit of the DFR proposal was 
rather original. While Snyder's space coordinates have discrete spectra so to 
induce a covariant analogue of lattice discretization, the DFR coordinates all
have continuous spectra, so that no limitations arise on the precision
of the localization \emph{in one coordinate}. Limitations arise instead on the
precision of simultaneous localization in two or more non commuting 
coordinates.

It is common folk lore\footnote{Apparently the first who made this remark 
(in a slight different form, namely revisiting the Heisenberg's microscope
\emph{gedankenexperiment}) 
was C.\ Alden Mead as early as 1959, 
but his paper underwent referee troubles  and was 
published in~1964~\cite{mead}. See the interesting letter of Mead to Physics 
Today \cite{mead2}.} 
that limitations on localization in the small should 
arise since, according to our understanding
of high energy physics, the localization process 
requires that a certain amount of energy
is transferred to the geometric background: the smaller the localization
region, the higher the energy density induced in the localization region. 
If the localization process reaches a 
sufficiently small length scale (typically the Planck 
length scale \(\lambda_P=
\sqrt{G\hbar c^{-3}}\)), a closed horizon might trap the region under 
observation, preventing any information to escape from it. 

This classical argument has been
invoked for example to claim that it is not 
possible to localize with a precision below the Planck length scale
(see e.g. \cite{dewitt,mead}). 
A moment's thought, however, would make it clear that such a statement is not
really supported by the above argument. Indeed, one might envisage to 
localize below the Planck length scale in one space dimension, at the cost
of sufficiently delocalizing in the remaining space dimensions. The resulting 
admissible localization region would be very large (compared with 
\(\lambda_P\)) and very thin; so thin to obtain 
localization much below the Planck length (in one direction), and so large
to keep the energy density sufficiently low to avoid black hole 
formation. It was precisely
this remark that led DFR to cast limitations on the admissible localization
regions in the form of uncertainty relations among the coordinates. In the
lack of more direct motivations, they decided to reproduce the path which,
long ago, led to canonical quantization; namely to find commutation 
relations inducing precisely the required uncertainty relations.

\subsection{The DFR Basic Model}
A heuristic analysis led DFR to postulate a very simple toy model of a flat,
fully covariant quantum spacetime, described by four quantum coordinates,
i.e.\ four selfadjoint operators $q^\mu$
on the infinite dimensional, separable
Hilbert space \(\mathfrak H\). Setting \(\lambda_P=1\) (in suitable units) 
and
\[
Q^{\mu\nu}=-\I[q^\mu,q^\nu],
\] 
the commutation relations are
\([q^\mu,Q^{\varrho\sigma}]=0\), or equivalently\footnote{More precisely,
equation \eqref{centralcom} gives the formal
relations \([q^\mu,Q^{\varrho\sigma}]=0\) 
the precise mathematical status of regular, strong commutation relations.}
\begin{equation}
\label{centralcom}
\exp(\I k_\mu q^\mu)\exp(\I h_\mu q^\mu)=\exp\Big(-\frac{\I }{2}k_\mu 
Q^{\mu\nu}h_\nu\Big)
\exp(\I(h+k)_\mu q^\mu),
\end{equation}
to be complemented with the statement 
that the joint spectral values \(\sigma^{\mu\nu}\) of the pairwise
commuting operators \(Q^{\mu\nu}\) define precisely the set \(\Sigma\) of
the antisymmetric matrices \(\sigma\) fulfilling\footnote{We recall that the
Hodge dual \(\ast\sigma\) of an antisymmetric 2-tensor \(\sigma\) is given by
\(
(\ast\sigma)_{\mu\nu}=(1/2)\varepsilon_{\mu\nu\lambda\varrho}
\sigma^{\lambda\varrho}.
\)}
\[\sigma_{\mu\nu}\sigma^{\mu\nu}=0,\quad
(\sigma^{\mu\nu}(\ast \sigma)_{\mu\nu})^2=16.
\]

The requirement that \([q^\mu,Q^{\varrho\sigma}]=0\) is a simplifying,
otherwise arbitrary ansatz; once this ansatz is accepted, the limitations
on the set \(\Sigma\) stem out of the DFR stability condition
of spacetime under localization, together with the 
quest for covariance.
We will not give the details; the interested reader is referred to the 
original paper. 

The coordinates are covariant: there is a unitary representation \(U\) of the 
Poincar\'e group \(\mathscr P\), such that
\[
U(\Lambda,a)q^\mu U(\Lambda,a)^{-1}={\Lambda^\mu}_\nu q^\nu+a^\mu,
\quad (\Lambda,a)\in\mathscr P.
\]

In strict analogy with Weyl quantization \cite{weyl}, 
one may consider the quantization
of an ordinary function \(f=f(x)\) of \(\mathbb R^4\) defined by
\begin{equation}
\label{quant}
f(q)=\int_{\mathbb R^4}dk\;\check f(k)e^{\I k q},
\end{equation}
where \(kq=k_\mu q^\mu\), and
\[
\check f(k)=\frac{1}{(2\pi)^4}\int_{\mathbb R^4}dx\;f(x)e^{-\I kx}.
\]

At this point, one might wish to follow the suggestion of Weyl (for example
like von Neumann did \cite{neumann}) and describe the operator product 
\(f_1(q)f_2(q)\) in terms of a suitable product of the
ordinary functions  \(f_1,f_2\). Unfortunately, the set of all the operators
of the form \(f(q)\) is not closed under operator products. It is then 
necessary to preliminarly 
extend the DFR quantization to a larger class of functions, 
namely the functions \(F=F(\sigma,x)\) of \(\Sigma\times\mathbb R^4\).
The full DFR quantization of a function of the form
\[
F(\sigma,x)=\sum_i g_i(\sigma)f_i(x)
\]
(a DFR symbol) is given by
\[
F(Q,q)=\sum_i g_i(Q)f_i(q);
\]
above, \(f_i(q)\) is understood as in \eqref{quant}, while \(g_i(Q)\) 
is the joint functional calculus of the pairwise commuting operators 
\(Q^{\mu\nu}\).

We may now define a covariant product $\star$ of two DFR symbols 
(the DFR twisted product) by requiring that
\[
F_1(Q,q)F_2(Q,q)=(F_1\star F_2)(Q,q);
\]
By standard computations, one easily finds
\begin{equation}\label{dfr_position}
\begin{split}
&(F_1\star F_2)(\sigma,x)=\\
&=\frac{1}{\pi^4}
\int\limits_{(\mathbb R^4)^2} dadb\;
F_1(\sigma,x+a)F_2(\sigma,x+b)\exp\big(-2\I
b_\mu(\sigma^{-1})^{\mu\nu}a_\nu\big).
\end{split}
\end{equation}
An asymptotic expansion of the (reduced) 
DFR product is widely known as the Moyal 
product. See the appendix for more details.

Following \cite{dfr}, we may define two maps, which, for self evident reasons, 
we will denote by \(\int_{\{q^0=t\}} d^3q\) and \(\int d^4q\), respectively:
\[
\int\limits_{\{q^0=t\}} d^3q\;F(Q,q)=
\int\limits_{\mathbb R^3}d^3\bx\;F(Q,(t,\bx)), \quad\quad 
\int d^4q\;F(Q,q)=\int\limits_{\mathbb R^4}d^4x\;F(Q,x).
\]
These maps are positive: for all  \(F\)'s and \(t\)'s,
\[
\int\limits_{\{q^0=t\}} d^3q\;F(Q,q)F(Q,q)^*\geqslant 0,
\quad\quad 
\int d^4q\;F(Q,q)F(Q,q)^*\geqslant 0.
\]
In particular, the positivity of the map \(\int_{\{q^0=t\}} d^3q\) is 
compatible with the uncertainty relations, since the latter allow for exact 
localization in \(q^0\), at the cost of total delocalization in the remaining
coordinates. Note also that, for any \(x\in\mathbb R^4\) fixed, the map
\(F(Q,q)\mapsto F(Q,x)\) is \emph{not} positive.

\subsection{Field Theory}
A (Wightman, say) quantum field \(\phi(x)\) on ordinary Minkowski spacetime
is a (generalized) function taking values (morally)
in the field algebra \(\mathfrak F\), 
namely it is (morally) in 
\(\mathcal C(\mathbb R^4,\mathfrak F)\simeq
\mathcal C(\mathbb R^4)\otimes\mathfrak F\). Here 
\(\mathcal C(\mathbb R^4)\) is the localization algebra.

Hence, it is natural to replace the classical localization algebra \(\mathcal
C(\mathbb R^4)\) with its quantized counterpart, the algebra \(\mathcal E\)
generated by the quantum coordinates \(q^\mu\). In this ``semiclassical''
model, it is natural to seek for quantum fields on quantum spacetime as 
elements of (morally) the algebra \(\mathcal E\otimes\mathfrak F\).
By analogy with the quantization of ordinary functions, DFR proposed the 
following quantization of the free Klein--Gordon field:
\[
\phi(q):=\int\limits_{\mathbb R^4} dk\;e^{\I kq}\otimes\check\phi(k).
\]
Then they made the following, fundamental remark. Let \(\mathscr 
H_0(\phi(x),\partial_\mu\phi(x))\) be the free Hamiltonian density. It is well
known that the full free Hamiltonian 
\[
H_0=\int d^3\bx\;\mathscr H_0(\phi(t,\bx),
\partial_\mu\phi(t,\bx))
\]
does not depend on the time \(t\). Then, it was shown in \cite{dfr} that
\[
\int\limits_{\{q^0=t\}}d^3q\;\mathscr H_0(\phi(q),\partial_\mu\phi(q))=H_0
\]  
(as a constant function of \(\sigma\)).
To put it in a more explicit way, with
\[
\mathscr H_0(\phi(x),\partial_\mu\phi(x))=
\frac{1}{2}{\boldsymbol :}\!
\left(\sum_\mu(\partial_\mu\phi)^2(x)+m^2\phi^2(x)\right)\!{\boldsymbol :},
\]
we have
\[
\begin{split}
&{\int\limits_{\{q^0=t\}}d^3q\;\mathscr H(\phi(q),
\partial_\mu\phi(q))=}\\
&=\frac{1}{2}\int\limits_{\mathbb R^3}d\bx\;
{\boldsymbol :}\!
\left(\sum_\mu((\partial_\mu\phi)\star(\partial_\mu\phi))(t,\bx)+
m^2(\phi\star\phi)(t,\bx)\right)\!{\boldsymbol :}=H_0
\end{split}
\]
(as a constant function of \(\sigma\)).

This remark\footnote{This result is independent from the well known 
\emph{general} fact that, for any two admissible functions \(f,g\), 
not necessarily solutions of the Klein--Gordon equation, then
 \(\int d^4x\;(f\star g)(x)\equiv\int d^4x\;(fg)(x)\). The latter result
is commonly taken as the starting point for a perturbative approach to 
nonlocal theories in the Euclidean setting, since it implies  
that the free action is unchanged by replacing the ordinary pointwise product
with some twisted product~\(\ast\). Note however that the Wick rotation of a 
twisted product is ill defined (namely no well defined~\(\ast\) may be 
reached by Wick rotating the DFR product~\(\star\)), 
so that at present (and to the best of the author's knowledge) 
the only known relation 
between Euclidean and Minkowskian ``twisted theories''
is a weak, indirect formal analogy.} 
is the starting point
for defining an effective perturbation theory on the
ordinary Minkowski quantum spacetime, in the so called interaction
representation; of course, this ansatz should be taken with
a grain of salt. Since the Hamiltonian is unaffected by the 
replacement of ordinary pointwise products with twisted products, 
we may consider
perturbations of the usual free fields. The only remaining difficulty
is the dependence on \(\sigma\); to get rid of it in view of an effective
theory on ordinary spacetime, we need to integrate it out by means
of some measure on \(\Sigma\). Unfortunately, there is no 
Lorentz invariant measure on \(\Sigma\); the best we can find is a rotation
invariant measure \(d\sigma\). Quite unfortunately, 
this will destroy the covariance of this class
of models under Lorentz boosts.

By analogy with the case of the local 
\(\phi^n\) interaction, we may consider the interaction Hamiltonian
\[
\begin{split}
H_I(t)&=\int\limits_{q^0=t}:\!\phi(q)^n\!:\;=\\
&=\int\limits_{\mathbb R^3}d\ba\;
\boldsymbol:\!(\phi\star\dotsm\star\phi)(Q,(t,\ba))\!\boldsymbol:,
\end{split}
\]
where \(\boldsymbol{:}\!AB\dotsm\boldsymbol{:}\) denotes the normal (i.e.\ Wick) ordering of operator
products. To obtain an effective interaction Hamiltonian on the classical
Minkowski spacetime, we integrate the \(\sigma\) dependence out, getting
\[
\begin{split}
H_I^{\text{eff}}(t)&=\\
&=\int\limits_{\Sigma} d\sigma\int\limits_{\mathbb R^3}d\ba\;
\boldsymbol:\!(\phi\star\dotsm\star\phi)(\sigma,(t,\ba))\!\boldsymbol:=\\
&=\int\limits_{\mathbb R^3}d\ba\int\limits_{\mathbb R^{4n}}
dx_1\dotsm dx_n\;W_{(t,\ba)}(x_1,\dotsc,x_n):\!\phi(x_1)\dotsm\phi(x_n)\!:
\end{split}
\]
for a suitable kernel\footnote{If \(C_\sigma\) is the kernel such that 
\[
(f_1\star\dotsm\star f_n)(\sigma,x)=\int\limits_{(\mathbb R^4)^n} dx_1\dotsm dx_n\;
C_\sigma(x-x_1,\dotsm,x-x_n)f_1(x_1)\dotsm f_n(x_n),
\] 
then 
\[
W_x(x_1,\dotsc,x_n)=
\int\limits_\Sigma d\sigma\;C_\sigma(x-x_1,\dotsc,(x-x_n).
\]} 
\(W_x\).
Following Dyson \cite{dyson}, the (formal) $S$-matrix is given by
\[
S=I+\sum_{N=1}^\infty\frac{(-\I g)^N}{N!}\int\limits_{\mathbb R^N} 
dt_1\dotsm dt_N\;
T[H_I^{\text{eff}}(t_1),\dotsc, H_I^{\text{eff}}(t_N)],
\]
where 
\[
\begin{split}
&T[H_I^{\text{eff}}(t_1),\dotsc, H_I^{\text{eff}}(t_N)]=\\
&=\sum_\pi H_I^{\text{eff}}(t_{\pi(1)})\dotsm H_I^{\text{eff}}(t_{\pi(N)})
\prod_{k=1}^{N-1}\theta(t_{\pi(k+1)}-t_{\pi(k)})
\end{split}
\]
is the product of the \(H_I^{\text{eff}}(t_j)\)'s taken in the order of 
decreasing times 
(the sum over \(\pi\) running over all permutations of \((1,\dotsc,N)\)).
Note that, contrary to the usual conventions, we wrote \(T[A,B,\dotsc]\)
instead of the traditional \(T[AB\dotsm]\).

As pointed out in \cite{dfr}, the time ordering acts on the overall
times \(t_1,\dotsc,t_n\) of the interactions Hamiltonians, not on the time 
arguments of the fields which appear in the definition of the interaction 
Hamiltonian. Since the perturbation theory
described above is built on top of a Hamiltonian model, the S-matrix of the 
DFR \(\phi^{\star n}\) interaction is (formally) unitary by construction.
More recent concerns about possible unitarity violation were the
consequence  of a too naive way of performing the time ordering prescription
(see \cite{unitarity}, and references therein; see also 
\cite{dorotesi,sibold}). The ultraviolet regularity of a \(\phi^{\star 3}\)
DFR model has recently been proved by Bahns \cite{bahns2},
under a weaker prescription for averaging over \(\sigma\). 

At this point it might be natural to convince ourselves that the trick of 
absorbing the time ordering into some analogue of the Stueckelberg--Feynman 
propagator  is not
possible any more. Indeed, in the local case the Stueckelberg--Feynman 
propagators allow us to 
consider one diagram, describing at once all possible arrangements of the time 
of the vertices \cite{stu,fey}; 
this feature might seem apparently lost in the present nonlocal 
case. We will see in the next section how, on the contrary, things are bound 
to go exactly the same way as in the local case.

Before closing this section, let us recall that 
different perturbative approaches,
which are equivalent in the local case, may well fail to be such
in the nonlocal case. As an example of this situation we mention
the noncommutative analogue of the Yang--Feldman equation proposed in
\cite{unitarity}, and developed in \cite{quasiplanar,dorotesi}. Finally, see
\cite{ultraviolet,gheratesi}
for a different generalization of the Wick product, based on optimal 
localization; the resulting model is free of ultraviolet divergences.

\section{Nonlocal Dyson Diagrams}
\label{sect_3}
In \cite{denk}, Denk and Schweda showed that the above mentioned concerns
were wrong; indeed, they were able to show that it was possible to absorb
the time ordering in the definition of a simple generalization of the
Stueckelberg--Feynman propagator, namely
\[
\mathscr{D}(x;\tau)=\frac{1}{\I}\left(\Delta_+(x)\theta(\tau)+
\Delta_+(-x)\theta(-\tau)\right).
\] 
When the theory is local, then the above general propagator (the 
``contractor'', according to Denk and Schweda) is always evaluated
at \(\tau=x^0\), in which case it reproduces the usual Stueckelberg--Feynman 
propagator:
\[
\mathscr D(x;x^0)=\Delta_{SF}(x).
\]

The original argument (which is casted in the case of non Wick ordered 
interactions, and worked out in the framework of the 
Gell--Mann \& Low formula) 
is rather involved: it relies on a clever, though very tricky 
manipulation consisting of Wick reducing
the \emph{ordinary} pointwise products of fields, and recombining the 
products of two--point functions and \(\theta\) functions by hand.  

There is, however, a profound reason why such a clever rearrangement of terms 
leads to the desired result: indeed, one can instead reproduce exactly
the same line of reasoning which can be found in any standard textbook
on local quantum field theory, since the Wick reduction 
of \emph{time
ordered} products of Wick monomials can be performed in the nonlocal 
setting considered here, too \cite{newrules}. In other words, the second Wick 
theorem is not 
local; this is so deeply true that even the original proof of Wick
does not rely on locality \cite{wick}. 

Let us see this in the case of the Dyson diagrams. Consider 
the \(N^{\text{th}}\) order contributions to the S-matrix:
\[
S^{(N)}=\frac{(-\I g)^N}{N!}\int\limits_{\mathbb R^N} 
dt_1\dotsm dt_N\; T[H_I^{\text{eff}}(t_1),\dotsc, H_I^{\text{eff}}(t_N)]
\]
We introduce the following shorthands:
\[
\ux_j=({x_j}_1,\dotsc,{x_j}_n)\in\mathbb R^{4n},
\quad d\ux_j=\prod_{k=1}^nd{x_j}_k,
\quad \phi^{(n)}(\ux_j)=\phi({x_j}_1)\dotsm\phi({x_j}_n),
\]
so that
\[
H_I^{\text{eff}}(t)=\int\limits_{\mathbb R^3} d\ba
\int\limits_{\mathbb R^{4n}}d\ux\;W_{(t,\ba)}
\boldsymbol:\!\phi^{(n)}(\ux)\!\boldsymbol:,
\]
and a short, standard computation gives\footnote{We rename 
\(t_j\) as \({a_j}^0\), so that \(\int_{\mathbb R} dt_j
\int_{\mathbb R^3} d\ba_j=\int_{\mathbb R^4}da\).}
\begin{equation}
\label{SN}
\begin{split}
S^{(N)}=&\frac{(-\I g)^N}{N!}
\int\limits_{(\mathbb R^{4})^N}da_1\dotsm da_N
\int\limits_{(\mathbb R^{4n})^N} d\ux_1\dotsm d\ux_N\;
W_{a_1}(\ux_1)\dotsm W_{a_N}(\ux_N)\times\\
&\times
T^{{a_1}^0,\dotsc,{a_N}^0}[
\boldsymbol:\!\phi^{(n)}(\ux_1)\!\boldsymbol:,\dotsc,
\boldsymbol:\!\phi^{(n)}(\ux_N)\!\boldsymbol:],
\end{split}
\end{equation}
where we introduced the following, natural notation:
\[
T^{\tau_1,\tau_2,\dotsc,\tau_k}[A_1,A_2,\dotsc,A_k]=
\sum_\pi A_{\pi(1)})\dotsm A_{\pi(k)}
\prod_{j=1}^{k-1}\theta\big(\tau_{\pi(j+1)}-\tau_{\pi(j)}\big),
\]
namely the product of the \(A_j\)'s is taken in the order of decreasing
\(\tau_j\)'s. Note that this definition may be given in general; there is 
no need for any a priori relation between the factors \(A_j\)
and the parameters \(\tau_j\) which we may wish to attach to those factors;
we will call the above a \emph{general} time ordered product, to highlight
this fact.

The key remark here is that the mechanism for the Wick reduction of a
general Time ordered product works as usual. The only difference with respect 
to the local case is that, here, we have to keep in mind that to each field
there corresponds a parameter driving its position in the 
time ordered product; this was implicit in the local case, where the time 
parameter corresponding to each field \(\phi(x^0,\bx)\) was precisely \(x^0\). 

In view of this remark, we need for a notation which indicates this 
correspondence explicitly, e.g.\
\begin{equation}\label{wick_aux}
\boldsymbol:\!
\overbrace{\phi({x_1}_1)\dotsm\phi({x_1}_{n})}^{{a_1}^0}
\overbrace{\phi({x_2}_1)\dotsm\phi({x_2}_{n})}^{{a_2}^0}
\dotsm
\overbrace{\phi({x_N}_1)\dotsm\phi({x_N}_{n})}^{{a_N}^0}
\!\boldsymbol:
\end{equation}
Wick contractions\footnote{Some reader might be surprised (I was surprised) 
to discover that the 
standard graphic notation for contractions is not due to Wick; it was
first introduced by Houriet and Kind in Helv.\ 
Phys.\ Acta {\bf 22}, 319 (1949). In his paper,
Wick politely complains for he was forced to abandon that very convenient
notation for typographical reasons. Hence, he writes e.g. 
\(:U^. VW^{..}X^{..}Y^. Z:\)
instead of \(\underwick{21}{:<1 U V <2 W >2 X >1 Y Z:}\).}, 
then, may be defined in the obvious way with respect to
the above correspondence:
\begin{gather*}
\boldsymbol:\!
\phi({x_1}_1)\dotsm
\underwick{2}{<1\phi({x_i}_u)
\cdot\cdot\cdot>1\phi({x_j}_v)}\cdot\cdot\cdot
\phi({x_N}_{n})\!\boldsymbol:
\quad=\\=\quad
\boldsymbol:\!
\phi({x_1}_1)\dotsm
\widehat{\phi({x_i}_u)}\dotsm
\widehat{\phi({x_j}_v)}\cdot\cdot\cdot
\phi({x_N}_{n})\!\boldsymbol:
\mathscr D({x_i}_u-{x_j}_v;{a_i}^0-{a_j}^0),
\end{gather*}
where a caret indicates omission. The reader is invited to read the beautiful
original paper of Wick \cite{wick}, to convince herself that the proof
does not rely on the requirement that the time associated to the field
is the same as the time argument of the field (see also \cite{newrules}
for a more detailed, ``non local'' discussion). Hence, we still may state the
second general Wick theorem, according to which 
\begin{quote}
\emph{The general 
time ordered product 
\[T^{{a_1}^0,\dotsc,{a_N}^0}[
\boldsymbol:\!\phi^{(n)}(\ux_1)\!\boldsymbol:,\dotsc,
\boldsymbol:\!\phi^{(n)}(\ux_N)\!\boldsymbol:]\] 
equals the sum of the
terms obtained by applying all possible choices of any number (including none) 
of allowed general contractions to \eqref{wick_aux}, where no
contraction is allowed, which involves two fields associated with the
same time parameter \({a_j}^0\).
}
\end{quote}
By applying the Wick reduction to \eqref{SN}, we obtain a certain sum of 
integrals, which we may label by means of Dyson diagrams, namely diagrams
consisting of \(N\) (= the order in perturbation theory)  
vertices, each of which is the origin of no more than \(n\) 
(possibly none) lines for a \(\phi^{\star n}\)
interaction; each line connects two distinct vertices 
(no loops, no external lines).  
This is possible because a Wick monomial 
\(\boldsymbol:\!\phi(x_1)\dotsm\phi(x_n)\!\boldsymbol:\) is totally symmetric
in the arguments \(x_1,\dotsc,x_n\), so that we may safely replace 
each kernel \(W_a(x_1,\dotsc,x_n)\) by its totally symmetric part in 
\eqref{SN}. To see how to proceed, let us consider the following 
second order contribution in the case of a \(\phi^{\star 3}\) interaction:
\begin{align*}
\frac{-g^2}{2}\int\limits_{\mathbb (R^4)^2}&dadb
\int\limits_{\mathbb (R^{4n})^2}d\ux d\uy\;
W_a(\ux)W_b(\uy)
\boldsymbol:\!
\underwick{23}{<1\phi(x_1)<2\phi(x_2)\phi(x_3)
\phi(y_1)>1\phi(y_2)>2\phi(y_3)}
\!\boldsymbol:\;=\\
=&\frac{-g^2}{2}\int\limits_{\mathbb (R^4)^2}dadb
\int\limits_{\mathbb (R^{4n})^2}d\ux d\uy\;
W_a(\ux)W_b(\uy)
\boldsymbol:\!
\phi(x_3)\phi(y_1)
\!\boldsymbol:\times\\
&\times
\mathscr D(x_1-y_2;a^0-b^0)\mathscr D(x_2-y_3;a^0-b^0)
\end{align*}
It is clear that, up to renaming the integration variables and using the
total symmetry of \(W_a,W_b\) and \(\boldsymbol:\!\phi(x_3)\phi(y_1)\!
\boldsymbol:\), the above integral is exactly the same that we would obtain
from the contribution
\[
\frac{-g^2}{2}\int\limits_{\mathbb (R^4)^2}dadb
\int\limits_{\mathbb (R^{4n})^2}d\ux d\uy\;
W_a(\ux)W_b(\uy)
\boldsymbol:\!
\underwick{23}{<1\phi(x_1)<2\phi(x_2)\phi(x_3)
>2\phi(y_1)>1\phi(y_2)\phi(y_3)}
\!\boldsymbol:
\]
Hence, the only informations which are needed to write it down 
are that there are two Wick monomials and two contractions between those
two monomials. To give this information, it is sufficient to draw a diagram
with two vertices (the two Wick monomials) and two lines connecting them
(the contractions). Of course, one also has to count the multiplicity of a 
diagram, namely the number of different contraction schemes that would lead to 
the same integral up to dummy integration variables.
\newpage
\section{Conclusions}
\label{sect_4}
We have seen that the usual diagrammatic expansion of the Dyson series
is not special to local interactions. Indeed, diagrams are a consequence 
of two non local tools, namely the Dyson perturbation series and the normal 
(Wick) ordering
of products of creation and annihilation parts. Both these tools are
completely unrelated to locality, hence this result should have not
come out as a surprise. Feynman diagrams require an additional tool,
the Gell--Mann \& Low formula \cite{gml}, which also has nothing to do with locality;
indeed, it was shown in \cite{newrules} that the dear old 
Feynman diagrams also arise
naturally in the reduction of the nonlocal Green functions of the DFR
perturbative model.

Actually, we did not introduce any really new argument, everything was
already virtually contained in the papers of Dyson, Wick, and Gell--Mann \& 
Low. The case of the usual local \(\phi^n\) interaction can be reobtained
as a special case, by setting 
\(W_a(x_1,\dotsc,x_n)=\prod_j\delta^{(4)}(x_j-a)\) in \eqref{SN}.

The unified treatment of both local and nonlocal
\(\phi^n\) interactions may be of some practical interest, since it allows to 
study the convergence of the large scale limit diagramwise. Moreover, it may 
allow for developing a renormalization scheme for nonlocal theories with
a strict correspondence between local and nonlocal subtractions. In the large 
scale limit, it might be natural to expect that nonlocal, possibly finite
subtractions converge to the infinite subtractions of the local renormalized 
theory. However, it should be kept in mind that a different point of view 
may well be taken. For example, in \cite{quasiplanar}, a 
different prescription for the admissible subtractions was investigated, 
in order to only select those subtractions which are divergent already
at the Planck length scale. 

We close with the following remark. 
Many years ago, Caianiello raised some concerns about what he regarded as
a too naive interpretation of Feynman diagrams as pictorial
representations of actual scattering processes (see \cite{caianiello}). 
He remarked that,
even if we were inclined to accept such a view prior to renormalization, 
some paradox might arise after implementing some subtraction prescription 
(e.g.\ in the fermionic case, strong interference among free field modes
belonging to different diagrams might produce violations of the exclusion
principle). Here, we found that, even 
prior to renormalization, diagrams seem to be nothing more than a graphic
representation of the CCR algebra combined with ordinary quantum mechanical
perturbation theory; it might be misleading to try to see more than that.

\newpage
\section*{Appendix. Twisted Products}
Actually, for both historical and technical reasons, the DFR twisted 
product was laid down in Fourier space\footnote{Twisted products first
arose in the framework of canonical quantization, in the late 1920's. 
Their use was first
advocated by Weyl \cite{weyl}, who however did not publish explicit 
equations; following Weyl's suggestion, von Neumann \cite{neumann} laid down
the twisted product in Fourier space (twisted convolution). 
The twisted product in position space first appeared (in the form of an 
asymptotic expansion) in a paper by Gr\"onewold; the integral form 
was first used by Baker and explicitly
written down by Pool. The first rigorous results on asymptotic expansions 
of twisted products  are probably due to Antonet, 
and a comprehensive investigation can
be found in \cite{varilly}, to which we also refer for the bibliographical 
coordinates missing in this footnote. The seminal work of 
Weyl and von Neumann inspired Wigner to define the so called Wigner 
transform; Wigner's work in turn led Moyal to define the so called 
Moyal bracket or sine--bracket 
\(\{f,g\}_\star=f\star g-g\star f\); the Moyal bracket then plaid a 
fundamental role
in a seminal paper by Bayen et al about geometric quantization of phase
manifolds. 
The covariant version
of the twisted product was first introduced by DFR in order to quantize 
the spacetime. For some strange reason, in
the current literature about QFT on noncommutative spacetime,
the DFR variant of the Weyl--von Neumann twisted product is widely known 
as the Moyal product.}.  
To avoid confusion, in this appendix
we will reserve the symbol \(\times\) to indicate the ordinary convolution
product, and \(\ttimes\) to indicate the twisted convolution
product (twisted product in Fourier space). Setting
\begin{align*}
(\varphi_1\ttimes\varphi_2)(\sigma,k)=\int\limits_{\mathbb R^4} 
dh\;\varphi_1(\sigma,h)\varphi_2(\sigma,k-h)
\exp\Big(\frac{\I}{2}k_\mu&\sigma^{\mu\nu}h_\nu\Big),\\ 
&\varphi_j(\sigma,\cdot)\in L^1(\mathbb R^4),
\end{align*}
one easily finds
\[
F_1\star F_2=\widehat{\check F_1\!\ttimes\! \check F_2},\quad\quad 
F_j(\sigma,\cdot)\in 
L^1(\mathbb R^4)\cap \widehat{L^1(\mathbb R^4)}.
\]

A (formal) asymptotic expansion of the product \(\star\) 
can be easily obtained by standard Fourier 
theory\footnote{With our conventions, we have 
\((\widehat{\partial_\mu f})(k)=-\I k_\mu\hat f(k)\) and
\(\hat f\times \hat g=(2\pi)^4\widehat{fg}\).}. 
Indeed, by replacing the exponential factor 
\(\exp[(\I/2)k\sigma h]=\exp[-(\I/2)h\sigma(k-h)]\) with
the corresponding exponential series and (formally) 
exchanging the sum and the integration, one obtains
\begin{eqnarray*}
\lefteqn{(F_1\star F_2)(\sigma,x)=}\\
&=& \int\limits_{\mathbb R^4} 
dk\;e^{ikx}\int\limits_{\mathbb R^4}
dh\;\check F_1(\sigma,h)\check F_2(\sigma,k-h)
\exp\Big\{\frac{\I}{2}\sigma^{\mu\nu}k_\mu(h-k)_\nu\Big\}\text{``\(=\)''}\\
&\text{``\(=\)''}&F_1(\sigma,x)F_2(\sigma,x)+\\
&&+\sum_{n=1}^\infty\frac{(-i/2)^n}{n!}
\sigma^{\mu_1\nu_1}\dotsm \sigma^{\mu_n\nu_n}
\big((\partial_{\mu_1}\dotsm \partial_{\mu_n}F_1)
\partial_{\nu_1}\dotsm \partial_{\nu_n}F_2\big)(\sigma,x).
\end{eqnarray*}

The above formal series gives a precise meaning to the more compact 
definition
\begin{equation}
\label{astw}
\mathscr{A}[F_1\star F_2](\sigma,x)=
\left(\exp\left\{-\frac{\I}{2}\sigma^{\mu\nu}
\partial_\mu\otimes\partial_\nu\right\}F_1\otimes F_2\right)(\sigma,x)
\end{equation}
of the asymptotic expansion \(\mathscr{A}[F_1\star F_2]\) of 
\(F_1\star F_2\); here, 
\(\otimes\) is a tensor product of functions defined fibrewise
over \(\sigma\) by 
\[
(F_1\otimes F_2)(\sigma,x,y)=F_1(\sigma, x)F_2(\sigma,y),
\]
and \(m\) is the fibrewise multiplication map
\[
m(F_1\otimes F_2)(\sigma,x)=F_1(\sigma,x)F_2(\sigma,x).
\]
Some authors take \eqref{astw} as their definition of twisted product.

This asymptotic expansion, however, may be rather misleading. Assume that,
for a fixed value of \(\sigma\), the functions \(F_j(\sigma,x)\) of \(x\)
are \(\mathcal C^\infty\) and have compact support, \(j=1,2\); 
moreover, assume that the supports are disjoint. Since derivatives cannot
enlarge the supports, we have 
\(\mathscr{A}[F_1\star F_2](\sigma,\cdot)\equiv 0\). In this precise
sense, the asymptotic expansion of the twisted product 
defines (in the sense of formal power series) a \emph{local}, non commutative 
product. This is rather unsatisfactory from the point of view of spacetime
quantization, since on general grounds we expect 
the quantum geometry to be non local.

Note also that, in the special case envisaged right above, 
there is no reason why we should have
\(F_1\star F_2\equiv 0\), and there are counterexamples, indeed. In other 
words, to naively rely on the above asymptotic expansion amounts to work with a
completely different algebra than that of ``true'' twisted products. This also 
shows that  the series may fail to converge (if it converges at all)
to the actual twisted product \(F_1\star F_2\). 

While for large classes of functions the asymptotic expansion truncated
at order \(n\) agrees with the twisted product  up to terms of order 
\(\lambda_P^{2(n+1)}\),
the issue of convergence is rather delicate, and there are very few general
results (see \cite{varilly}, and references therein). Certainly, the 
asymptotic expansion converges to (the Antonet extension of) 
\(F_1\star F_2\) if \(\check F_1,\check F_2\) have compact supports; 
in this case, \(F_1,F_2\) are real--analytic, i.e. they are nonlocal
in the sense that they cannot be deformed locally while preserving 
analiticity. In a sense, we may equivalently 
describe the nonlocality of quantized spacetime using
either \eqref{dfr_position} as a nonlocal product
of local objects (\(L^1\) functions),
or \eqref{astw} as a local product of 
nonlocal objects (analytic functions). As a (formal) product
of smooth functions, \eqref{astw} is irremediably local, and 
should be dismissed.


\begin{thebibliography}{99.}
\bibitem{newrules} G.\ Piacitelli,
\emph{Non local theories: New rules for old diagrams},
JHEP {\bf 0408} 031 (2004) 
\href{http://www.arXiv.org/hep-th/0403055}{[arXiv:hep-th/0403055]}.


\bibitem{dfr} 
S.\ Doplicher, K.\ Fredenhagen, and J.\ E.\ Roberts,
\emph{The quantum structure of spacetime at the Planck scale and quantum 
fields}, Commun.\ Math.\ Phys.\ {\bf 172}, 187--220 (1995)
\href{http://www.arXiv.org/hep-th/0303037}{[arXiv:hep-th/0303037]}.




\bibitem{snyder}
  H.\ S.\ Snyder, \emph{Quantized Space-Time},
  \href{http://prola.aps.org/abstract/PR/v71/i1/p38_1}%
{Phys.\ Rev.\  {\bf 71} 38 (1947)}.


\bibitem{mead}
C. A. Mead, Phys. Rev. \emph{Possible Connection Between Gravitation and 
Fundamental Length}, {\bf 135}, B849--B862 (1964).

\bibitem{mead2} C.\ A.\ Mead, \emph{Walking the Planck Length through History},
letter to the editor, 
\href{http://www.physicstoday.org/vol-54/iss-11/p15.html}%
{Phys. Today, {\bf 54}, N.\ 11 (2001)}.

\bibitem{dewitt} B.\ DeWitt, \emph{Quantizing Geometry}, 1962. In:
L. Witten (ed.),  \emph{Gravitation: An Introduction to Current Research}
New York, Wiley 1962.)


\bibitem{weyl} 
H.\ Weyl, \emph{Gruppentheorie und Quantenmechanik}, Hirzel, Leipzig 1928.

\bibitem{neumann}
J.\ von Neumann, \emph{Die Eindeutigkeit der Schr\"odingerschen Operatoren},
Math.\ Ann.\ {\bf 104} 570 (1931).

\bibitem{dyson}
F.\ J.\ Dyson,
\emph{The Radiation Theories Of Tomonaga, Schwinger, And Feynman},\\
\href{http://prola.aps.org/abstract/PR/v75/i3/p486_1}{Phys.\ Rev.\  {\bf 75} 486--502 (1949)}.




\bibitem{unitarity} 
D.\ Bahns, S.\ Doplicher, K.\ Fredenhagen, and G.\ Piacitelli,
\emph{On the Unitarity Problem in space/time noncommutative theories}
\href{http://dx.doi.org/10.1016/S0370-2693(02)01563-0}%
{Phys.\ Lett.\ B {\bf 533} 178-- (2002)}
\href{http://www.arxiv.org/hep-th/0201222}{[arXiv:hep-th/0201222]}. 

\bibitem{dorotesi} D.\ Bahns, \emph{Perturbative Methods on the Noncommutative
Minkowski Space}, Doctoral thesis, 
\href{http://www-library.desy.de/preparch/desy/thesis/desy-thesis-04-004.ps.gz}{DESY-THESIS-04-004}. Hamburg (2003).

\bibitem{sibold}
Y.~Liao and K.~Sibold,
\emph{Time-ordered perturbation theory on noncommutative spacetime: Basic
rules}
\href{http://www.edpsciences.org/articles/epjc/abs/2002/12/100520469/100520469.html}{Eur.\ Phys.\ J.\ C {\bf 25} 469 (2002)}
\href{http://www.arXiv.org/hep-th/0205269}{[arXiv:hep-th/0205269]}. 
\emph{Time-ordered perturbation theory on noncommutative spacetime. II.
Unitarity},
\href{http://www.edpsciences.org/articles/epjc/abs/2002/12/100520479/100520479.html}{Eur.\ Phys.\ J.\ C {\bf 25} 479 (2002)}
\href{http://www.arXiv.org/hep-th/0206011}{[arXiv:hep-th/0206011]}.



\bibitem{bahns2} D. Bahns, 
\emph{Ultraviolet Finiteness of the Averaged Hamiltonian on the Noncommutative Minkowski Space.}, to be published 
\href{http://www.arXiv.org/hep-th/0405224}{[arXiv:hep-th/0405224]}.


\bibitem{stu} D.\ Rivier and E.\ C.\ G.\ Stueckelberg,
\emph{A Convergent Expression for the Magnetic Moment of the Neutron},
\href{http://prola.aps.org/abstract/PR/v74/i2/p218_1}{letter to the Editor of 
Phys.\ Rev.\ {\bf 74}, 218 (1948)}, and references 
2,3,4 therein. 

 \bibitem{fey} R.\ P.\ Feynman,
\emph{Space-Time Approach To Quantum Electrodynamics},\\
\href{http://prola.aps.org/abstract/PR/v76/i6/p769_1}{Phys.\ Rev.\  {\bf 76} 
769--789 (1949)}. 



\bibitem{quasiplanar}BDFP quasiplanar
D.\ Bahns, S.\ Doplicher, K.\ Fredenhagen, and G.\ Piacitelli,
\emph{Field theory on noncommutative spacetimes: Quasiplanar Wick products},
  Phys.\ Rev.\ D {\bf 71}, 25022 (2005)
 \href{http://www.arXiv.org/hep-th/0408204]}{[arXiv:hep-th/0408204]}.




\bibitem{ultraviolet} 
D.\ Bahns, S.\ Doplicher, K.\ Fredenhagen, and G.\ Piacitelli,
\emph{Ultraviolet finite quantum field theory on quantum spacetime},
\href{http://www.springelink.de/link.asp?id=r9kgg5pptvj8fhp9}%
{Commun.\ Math.\ Phys.\  {\bf 237}, 221--241 (2003)}
\href{http://www.arXiv.org/hep-th/0301100}{[arXiv:hep-th/0301100]}.

\bibitem{gheratesi} G.\ Piacitelli, \emph{Normal ordering of Operator Products
on Noncommutative Space Time and Quantum Field Theory}, doctoral thesis, 2001.

\bibitem{denk} 
S.\ Denk and M.\ Schweda,
\emph{Time ordered perturbation theory for non-local interactions:  
Applications
to NCQFT},
\href{http://dx.doi.org/10.1088/1126-6708/2003/09/032}%
{JHEP {\bf 309}, 32 (2003)} 
\href{http://www.arXiv.org/hep-th/0306101}{[arXiv:hep-th/0306101]}.


\bibitem{wick}
G.\ C.\ Wick, \emph{The Evaluation of the Collision Matrix},\\
\href{http://prola.aps.org/abstract/PR/v80/i2/p268_1}{Phys.\ Rev.\ {\bf 80}, 
268--272 (1950)}.

\bibitem{gml} M.\ Gell-Mann and F. Low, \emph{Bound States in Quantum Field
Theory},\\
\href{http://prola.aps.org/abstract/PR/v84/i2/p350_1}{Phys.\ Rev.\ {\bf 84} 350--354 (1951)}.

\bibitem{caianiello} E.\ R.\  
Caianiello, \emph{Combinatorics and Renormalization
in quantum Field Theory}, W.\ A.\ Benjamin, Reading, Mass., 1973.



\bibitem{varilly} 
R.~Estrada, J.~M.~Gracia-Bondia, and J.~C.~Varilly,
\emph{On Asymptotic Expansions Of Twisted Products},
 J.\ Math.\ Phys.\  {\bf 30} 2789--2796 (1989).
\end{thebibliography}
 \end{document}